\documentclass[twocolumn ,prl]{revtex4}

\usepackage{graphicx}
\usepackage{amssymb}

\def\be{\begin{equation}}
\def\ee{\end{equation}}
\def\ben{\begin{equation} \nonumber}
\def\een{\end{equation}}
\def\ban{\begin{eqnarray*}}
\def\ean{\end{eqnarray*}}
\def\ba{\begin{eqnarray}}
\def\ea{\end{eqnarray}}
\def\({\left(}
\def\){\right)}

\def\half{{1\over2}}

\begin{document}

\title{The Gelaton Scenario:\\ Equilateral non-Gaussianity from multi-field dynamics}
\author{Andrew J. Tolley}
\email{atolley@perimeterinstitute.ca}
\author{Mark Wyman}
\email{mwyman@perimeterinstitute.ca}
\affiliation{Perimeter Institute for Theoretical Physics \\
	 31 Caroline St. N \\
	 Waterloo, ON  N2L 2Y5,
	 Canada\\}
\preprint{pi-cosmo-128}
%\date{}                                           % Activate to display a given date or no date

\begin{abstract}
The distinctive features of single field inflationary models with non-minimal kinetic terms,
like Dirac-Born-Infeld and $k$-inflation, can be captured by more familiar multiple field inflationary systems of the type that typically arise in low energy supergravity models. At least one heavy field, which we call the gelaton, has an effective potential which depends on the kinetic energy of the inflaton. Integrating out the gelaton gives rise to an effectively single field system for which the speed of sound for the adiabatic fluctuations is reduced, generating potentially observable equilateral non-Gaussianity, while causing negligible isocurvature fluctuations. This mechanism is only active if there is a relatively tight coupling between the gelaton and the inflaton. Requiring that the inflaton-gelaton system remains weakly coupled puts an upper limit on the gelaton mass. This approach gives a potentially UV-completable framework for describing large classes of $k$-inflationary behavior.

 \end{abstract}
\maketitle

Were the primordial perturbations that seeded structure formation purely Gaussian, or did they include a significant non-Gaussian
component? Cosmological observations will soon answer this question \cite{Komatsu:2009kd}.
Measuring a departure from Gaussianity would revolutionize our understanding of the inflationary epoch.
There are several ways in which non-Gaussianity is thought likely to arise.
These are named after the shapes of the triangles in momentum-space at which their amplitudes peak: equilateral, local/squeezed, orthogonal, and folded/flattened.
Discovering non-Gaussianity of a particular class would give us remarkable insight into the physics of inflation. An equilateral signal might arise from a self-interacting scalar field with a non-minimal kinetic term/higher dimension operators, like $k$-inflation \cite{ArmendarizPicon:1999rj} or Dirac-Born-Infeld (DBI) inflation \cite{Alishahiha:2004eh}; or from particle production during inflation \cite{Green:2009ds}. A local/squeezed configuration would
indicate the presence of multiple inflationary \cite{lyth/ungarelli/wands:2003} (or ekyprotic \cite{Buchbinder:2007at}) fields. A folded/flattened shape would only come from an excited
initial vacuum state \cite{Chen:2006nt,tolley:2008}. An orthogonal shape was identified in \cite{Senatore:2009gt}, arising in single field models. More generally we may expect any admixture of these different contributions.

In the following, 
we unveil a novel situation that does not  fit into the standard lore: a multiple-field
inflationary scenario that will exhibit equilateral-type non-Gaussianity by mimicking the dynamics of $k$-inflation or DBI 
while possessing minimal local-type non-Gaussianity.

Equilateral non-Gaussianity is uniquely tied to the departure of the speed of sound for adiabatic perturbations from the speed of light \cite{Cheung:2007st}. This can only happen in models for which the kinetic term is modified. In effective field theory language, a canonical single field kinetic term is dimension 4. Any modification requires the addition of irrelevant, or higher dimension, operators. This is made explicit in the now familiar $k$-inflation models, of which DBI inflation is the most well-motivated realization. The DBI case is special: the square root form of the action is protected by a non-linearly realized form of higher-dimensional Lorentz or Anti-de Sitter symmetry. This comes from its origin as the position modulus of a brane in higher dimensions.

There is another, perhaps more familiar, case in which there are large irrelevant operators modifying the kinetic term where quantum corrections may be under control. A good example is the bosonic truncation of low energy string/supergravity models. Setting to zero any gauge and form fields, we have
\ba
S & = & \int d^4 x \sqrt{-g} \left[ \frac{M_P^2}{2}R - \frac{1}{2}
 G_{IJ}\partial \phi^I \partial \phi^J   - V (\phi^I) \right]. \nonumber
\ea
The key novel feature of such a multi-field model is that it can have a non-trivial `metric on field space', $G_{IJ}$. In supergravity models, this metric is K\"ahler, and the theory is specified by its K\"ahler form and superpotential. Now, since only a canonically coupled kinetic term is dimension 4, any departure of $G_{IJ}$ from $\delta_{IJ}$ not equivalent to a field redefinition corresponds to a regime in which irrelevant operators are important. If the field space metric is non-Cartesian and under control over a field range of order the radius of curvature, then we can expect that the non-minimal form of the metric will create an order unity modification of the dynamics of the light fields.

Practically, this means that we can construct a viable inflationary mechanism in multi-field models that will give rise to large equilateral non-Gaussianity, the origin of which will be the field space metric $G_{IJ}$.

{\bf Turning down the speed of sound:}
The characteristic energy scale of inflation is set by the Hubble scale $H$. Light fields that have a mass which are comparable or lighter than this scale will generate long wavelength fluctuations, whereas heavy fields for which $m \gg H$ will oscillate and redshift. If we are only interested in looking at the evolution of the system as modes cross the horizon and at even longer wavelengths, it is permissible to integrate out the heavy fields. This gives rise to an effective theory of the light fields. Typically, the effect of the heavy fields is felt through small irrelevant operators which modify the light field Lagrangian, and are suppressed by the mass of the heavy field. An exception can occur if the heavy field is not in its vacuum state. The fluctuations of a massive field redshift like matter with the expansion, and the vacuum expectation value will come to rest at the minimum of its effective potential.
%However, if the kinetic term depends on the heavy field through $G_{IJ}$, its effective potential includes a contribution from the kinetic energy of the light fields. 
However, because of the $G_{IJ}$ coupling, the effective potential of the heavy field depends on the kinetic energy of the light fields. This means the heavy field gets forced into a coherent state whose expectation value is related to the kinetic energy of the light field. The more massive the heavy field, the stronger we must make the coupling with the light field to keep the heavy field away from its naive minimum. However, this coupling must not be made too strong. If it is, the entire effective field theory will break down. 

This prompts the following definition:
A {\it gelaton}, is a heavy field, $m>H$, which is sufficiently {\it strongly coupled} to the light fields that it creates an order unity modification of the light field dynamics, but is sufficiently {\it weakly coupled} that quantum corrections are under control.
The sense of the definition is that the gelaton gels, or sticks, to the light field, getting dragged along by it and in turn altering the dynamics of the light field.

To realize this, consider a well studied two field system:
\ba
\label{actionnc}
S & = & \int d^4 x \sqrt{-g} \left[ \frac{M_P^2}{2}R - \frac{1}{2}
(\partial_\mu\phi)(\partial^\mu\phi) -  \right . \nonumber\\
 && \left . \frac{e^{2b(\phi)}}{2}
(\partial_\mu\chi)(\partial^\mu\chi) - V (\phi, \chi) \right].
\ea
In the following we will make use of the notation and many of the general results calculated in refs.  \cite{DiMarco:2002eb,Lalak:2007vi}. We will take  $\chi$ to be the inflaton and $\phi$ to be the gelaton. If the effective mass of the gelaton is much larger than $H$, it will get frozen at the minimum of its effective potential, $\phi_o$, which is determined by the equation
\be
\label{con}
V_{,\phi}(\phi_o,\chi)-2b_{,\phi}(\phi_o) e^{2b(\phi_o)} X =0.
\ee
The solution is $\phi_o(\chi,X)$, where $X=-\frac{1}{2}(\partial \chi)^2$  and we take, {\it e.g.}, $V_{,\phi} \equiv \partial V / \partial \phi$. 
Substituting back into the action, and neglecting the kinetic term for $\phi$, we obtain
\be
S  =  \int d^4 x \sqrt{-g} \left[ \frac{M_P^2}{2}R +p(X,\chi)+\dots \right],
\ee
with 
\be
\label{con2}
p(X,\chi)=e^{2b(\phi_o(\chi,X))} X- V \left(\phi_o(\chi,X), \chi \right).
\ee
The resulting low-energy effective theory is thus classically equivalent, at leading order, to a $p(X,\chi)$ model, i.e., $k$-inflation or DBI-inflation \footnote{The Thomas-Fermi approximation of \cite{Bilic:2008pe} uses a similar technique.} . This is our principal result. In what follows, we demonstrate that it is a consistent effective theory. Quantum mechanically things are more subtle: we cannot switch off the quantum fluctuations in the $\phi$ field. These fluctuations, which increase with the mass of the gelaton, will tend to push the system away from its minimum. Thus there is no decoupling limit here obtained in the limit $m_{\rm gelaton} \rightarrow \infty$, as we shall see below.

Before continuing, we should demonstrate that such a model can still inflate, given a suitable choice of potential. This is straightforward to do.  A simple route is to follow the reasoning of Ref.~\cite{Tolley:2007nq}: power law inflationary solutions will arise whenever the action admits a scaling symmetry. This will be the case whenever $V(\phi,\chi)=e^{-c\chi}V(\phi)$. On integrating out the gelaton field, we obtain a model of the form considered in \cite{Piazza:2004df}: $p(X,\chi)= e^{-c\chi} F\left (e^{c\chi}X \right)$, which clearly preserves the scaling symmetry. If the gelaton mass is lighter than $H$, we recover a more familiar two field inflationary system with isocurvature modes, as considered in \cite{Tolley:2007nq}.

To understand the dynamics of the system, it is helpful to separate the {\it dressed} light degrees of freedom from the heavy oscillating degrees of freedom.In multi-field inflationary systems, one usually redefines the field perturbations into adiabatic and isocurvature/entropy modes. In the present case this decomposition is less useful. Since the position of the gelaton is tied to the field value and kinetic energy of the inflaton, the relevant transformation is actually a  canonical transformation. The full details of this canonical transformation are somewhat involved~\cite{prep}. Consequently, we will give a simplified argument here.

Following the standard procedure (e.g. \cite{DiMarco:2002eb,Lalak:2007vi}), we can define gauge invariant scalar perturbations $Q_{\phi}=\delta \phi+H^{-1}\dot{\phi}\psi$ and $Q_{\chi}=\delta \chi+H^{-1}\dot{\chi}\psi$, where $\psi$ is the spatial metric perturbation. We then perturb the action to second order, solve the constraints and substitute back in. We find the following two field system
\ba
\label{pertaction} &&S_{(2)} =\int d^4 x \frac{a^3}{2} \( \dot{Q_{\phi}}^2+e^{2b} \dot{Q_{\chi}}^2 -\(\frac{k^2}{a^2}+C_{\phi\phi}\)Q_{\phi}^2 \right .\\
&&\left . -\(\frac{k^2}{a^2}+C_{\chi\chi}\)e^{2b}Q_{\chi}^2 
 + 4b_{,\phi}e^{2b} \dot{\chi} \dot{Q_{\chi}} Q_{\phi} -2C_{\phi\chi}Q_{\chi}Q_{\phi} \). \nonumber
\ea
Here, $C_{\phi \phi}$, $C_{\phi \chi}$, and $C_{\chi\chi}$ are background-dependent coefficients, which are the mass matrix for the two perturbations as defined in \cite{Lalak:2007vi}. 
The effective mass of the gelaton perturbations is $m^2_{\rm gelaton}=C_{\phi\phi}  \gg H^2$. At first glance this action seems to describe a system of two fields each propagating with speed of sound $c_s=1$. At low energies however the coupling $Q_{\phi} \dot{Q}_{\chi}$ modifies the effective speed of sound, an effect which only arises when the field space metric is not Cartesian. The significance of this type of interaction was also considered in  \cite{Tolley:2007nq}.

The minimum of the effective potential for $Q_{\phi}$ is at
\be
\( m^2_{\rm gelaton}+\frac{k^2}{a^2} \)Q_{\phi}\sim 2 e^{2b}b_{,\phi}\dot{\chi} \dot{Q}_{\chi}-C_{\phi\chi} Q_{\chi}.
\ee
This equation describes the two field description of the dressed light degrees of freedom. In other words, it isolates the light component of $Q_{\phi}$. In general we can decompose $Q_{\phi}$ into its heavy and light parts via $Q_{\phi}=\hat{Q}_{\phi}-A Q_{\chi}+B \dot{Q}_{\chi}$, where 
\ba 
&&\( m^2_{\rm gelaton}+\frac{k^2}{a^2} \) A =C_{\phi\chi} \\
&& \( m^2_{\rm gelaton}+\frac{k^2}{a^2} \) B=2 e^{2b}b_{,\phi}\dot{\chi}.
\ea
Substituting in this redefinition, we obtain
\ba
\label{gelact}
 &&S_{(2)} =\int d^4 x \frac{a^3}{2} \left[ \dot{\hat{Q}_{\phi}}^2 -\(\frac{k^2}{a^2}+m^2_{\rm gelaton}\)\hat{Q}_{\phi}^2 \right . \nonumber \\
&& +e^{2b} \dot{Q_{\chi}}^2-\(\frac{k^2}{a^2}+C_{\chi\chi}\)e^{2b}Q_{\chi}^2 
  \\
 &&\left. +\(\frac{k^2}{a^2}+m^2_{\rm gelaton}\)\(AQ_{\chi}-B \dot{Q}_{\chi} \)^2 \right]+S_{(2)}^{\rm int} , \nonumber
\ea
where we have defined
\be
S_{(2)}^{\rm int}= \int d^4 x \frac{a^3}{2} \left[ \( \dot{\hat{Q}_{\phi}} -\dot{(AQ_{\chi})}+\dot{(B\dot{Q_{\chi}})}\) ^2-\dot{\hat{Q}_{\phi}}^2 \right].
\ee
The appearance of higher derivative terms in $S^{\rm int}$ is an artifact of our simple approach. They can be avoided by replacing the field redefinitions with a full canonical transformation~\cite{prep}. However, as long as $S^{\rm int}$ is small, the above argument is sufficient.
The key point is that in the gelaton regime, the interaction term $S_{(2)}^{\rm int}$ is suppressed by powers of $H^2/m^2_{\rm gelaton}$ and indeed may be treated as small.
On neglecting this term in Eqn. \ref{gelact}, we see that this system clearly separates into two modes, a heavy mode $\hat{Q}_{\phi}$ of mass $m_{\rm gelaton}$ and a dressed light degree of freedom $Q_{\chi}$.  If $H B \gg A$, then, at leading order,
\ba
 &&S_{(2)} \sim \int d^4 x \frac{a^3}{2} \left[ \dot{\hat{Q}_{\phi}}^2 -\(\frac{k^2}{a^2}+m^2_{\rm gelaton}\)\hat{Q}_{\phi}^2 + \right .   \\
&& \left.e^{2b} \dot{Q_{\chi}}^2-\(\frac{k^2}{a^2}+C_{\chi\chi}\)e^{2b}Q_{\chi}^2 +\(\frac{k^2}{a^2}+m^2_{\rm gelaton}\)B^2 \dot{Q}_{\chi}^2 \right].
  \nonumber 
\ea
The additional kinetic term from the inflaton fluctuations changes the effective speed of sound.
In general this effective speed of sound is $k$ dependent, but in the low energy regime $k/a\ll m_{\rm gelaton}$ it reduces to
\be
\label{cs}
c_s^2 = \( 1+\frac{4e^{2b}b_{,\phi}^2 \dot{\chi}^2}{m^2_{\rm gelaton}}\)^{-1}=\frac{ \half \frac{V_{\phi \phi}}{b_\phi V_\phi} - \half \frac{b_{\phi \phi}}{b_\phi^2} - 1}{ \half \frac{V_{\phi \phi}}{b_\phi V_\phi} - \half \frac{b_{\phi \phi}}{b_\phi^2}+1} \, .
\ee
Here, since $\dot{\phi} \sim 0$, we have (from eq. (19) in \cite{Lalak:2007vi})
$C_{\phi \phi} = m^2_{\rm gelaton} \simeq -2 V_\phi b_\phi - (V_\phi b_{\phi \phi}/b_\phi) + V_{\phi \phi}$.
As expected, the sound speed departs from unity whenever $b(\phi)$ is not a constant.

We should perhaps say one further word about the  low energy regime ($k/a\ll m_{\rm gelaton}$) in which we are working. It is only in this regime that 
the gelaton can safely be integrated out, bringing about the new dressed dynamics of the lighter field. In the high energy limit, both fields have usual 
dynamics and the sound speeds for each return to unity. 

{\bf Hyperbolic manifolds:} A field space metric that often arises in the context of supergravity theories are manifolds which are locally equivalent to hyperbolic spaces; i.e., $b(\phi)=g\phi/M_P$. In this case the above formula simplifies to 
$c_s^2 = \( 1-\frac{2gV_{,\phi}}{M_P V_{,\phi\phi}}\)\(  1+\frac{2gV_{,\phi}}{M_P V_{,\phi\phi}} \)^{-1}$. Using Eqs. (\ref{con}) and (\ref{con2}) it is straightforward to show that this expression is equivalent to the usual $k$-inflationary expression, $c_s^2=p_{,X}/\rho_{,X}$.
Using these relations, we can reproduce, for instance, the dynamics of DBI inflation,
whose Lagrangian is $p(X) = T(\chi)\sqrt{1 -2 X /T(\chi)} - T(\chi) + W(\chi)$. All that is required is this potential:
\be
V_{\rm{DBI}}(\phi, \chi)  = T(\chi) \cosh(2g\phi/M_P) - T(\chi) + W(\chi).
\ee
The effective gelaton mass is determined by the one free parameter $g$, $m^2_{\rm gelaton}= 4g^2M_P^{-2}T(\chi)\exp(-2g\phi/M_P)$.

The previous procedure is easily generalizable to the nonlinear level. Starting from the full action~(\ref{actionnc}), we may separate the field $\phi$ into its light and heavy parts as $\phi=\phi_o(X,\chi)+\hat{\phi}$. Expanding to cubic order in $\hat{\phi}$, we find
\ba
&& S= \int d^4 x \sqrt{-g} \left[ \frac{M_P^2}{2}R +p(X,\chi) \right. \\
&& -\frac{1}{2}(\partial \hat{\phi})^2 -\frac{1}{2}\(  V_{,\phi\phi}(\phi_o,\chi)-C_2(\phi_o)X\) \hat{\phi}^2 \nonumber \\ \nonumber
&& \left. -\frac{1}{6}\( V_{,\phi\phi\phi}(\phi_o,\chi)-C_3(\phi_o)X\) \hat{\phi}^3 +\dots\right]+S^{\rm int}
\ea
where $C_{n}=\partial_{\phi}^n \( e^{2b(\phi)}\)$.
As before, $S^{\rm int}$ is an interaction suppressed by the gelaton mass that we can neglect at low energies
\be
S^{\rm int}=\int d^4 x \sqrt{-g} \frac{1}{2}\((\partial \phi_o)^2+2\partial \phi_o \partial \hat{\phi} \).
\ee

{\bf Regime of validity:} The expression for the sound speed, Eq.~(\ref{cs}), makes the conditions necessary for a gelaton phase clear. As the mass squared, measured by $C_{\phi\phi}$, increases, the coupling between the gelaton and the inflaton measured by $b_{,\phi}$ ({\it i.e.} $g$ in the hyperbolic case) must increase accordingly for $c_s$ to depart significantly from unity. By making this coupling large we have effectively resummed an infinite set of irrelevant (gelaton mass suppressed) operators
\be
c_s^2 \sim 1-\(\frac{4e^{2b}b_{,\phi}^2 \dot{\chi}^2}{m^2_{\rm gelaton}} \)+\(\frac{4e^{2b}b_{,\phi}^2 \dot{\chi}^2}{m^2_{\rm gelaton}} \)^2+\dots
\ee
The crucial question is whether this resummation is stable under quantum corrections. This question has two parts: (a) is it technically natural? and (b) will perturbation theory remain weakly coupled, {\it i.e.} are we below the cutoff of the effective theory?

The first question is obviously more model dependent. It depends on the specific multi-field model under consideration. For instance, in the case of models which are embedded in the context of supergravity theories, we should expect better control over the field space metric. At least one relevant contribution is the loop correction from the gelaton field. We can get this using a simple generalization of the Coleman-Weinberg argument. On integrating out the gelaton but including the one-loop contribution, we find an effective single field $p(X,\chi)$ model with 
\ba
&& p_{\rm eff}(X,\chi)=e^{2b(\phi_o(\chi,X))} X- V \left(\phi_o(\chi,X), \chi \right) \\ \nonumber 
&&-\frac{1}{64\pi^2}\( V_{,\phi\phi}(\phi_o,\chi)-C_2(\phi_o)X\)^2 \ln \(\frac{V_{,\phi\phi}-C_2X}{\mu^2} \),
\ea
for some fiducial renormalization scale $\mu$. In a supergravity model we may expect similar order contributions from fermion loops. Here both the kinetic and potential terms get renormalized. As long as the gelaton mass is not too large, it gives no naturalness problem. Had we chosen to use an explicit cutoff $\Lambda$, we would also expect contributions of the form $\Lambda^4$ -- which would just renormalize the cosmological constant -- and at next order of the form $m^2_{\rm gelaton} \Lambda^2$. These terms give potentially large renormalizations of the kinetic term unless the cutoff is very low. However, since these contributions are clearly prescription dependent, we content ourselves with stability with respect to the running contribution. 

It is more important to consider the regime of validity of the effective theory. From the arguments of \cite{Cheung:2007st} and \cite{Leblond:2008gg}, we anticipate that the single field effective theory must break down at energy scales $\Lambda\gg H$ for which
\be
\label{cutoff}
\Lambda^4 \sim \( 16\pi^2M_{P}^2 |\dot{H}|c_s^5\) \(1-c_s^2\)^{-1}.
\ee
Since the cutoff for the single field effective theory is $m_{\rm gelaton}$, there must be an upper bound on the gelaton mass, $m_{\rm gelaton} \le \Lambda$.

This bound was derived using the inflaton perturbations, but a similar bound can be derived more directly by looking at the scale of the gelaton interactions. For example, in the DBI case, the gelaton mass $m_{\rm gelaton}^2 \sim g^2 M_P^{-2}c_s T(\chi)$.
Quantum effects cause $\phi$ to fluctuate away from the minimum of the effective potential by an amount $\Delta\hat{\phi} \sim m_{\rm gelaton}$ .
The coupling between the two fields is set by the exponential $e^{g\hat{\phi}/M_P}$. In light of this, we require $g \, m_{\rm gelaton} \le M_{P}$
for the approximation to be self-consistent. Rewriting this bound in terms of more physical quantities, we find that the gelaton mass must satisfy $H^4 \ll m^4_{\rm gelaton} \le c_s T$.
The condition Eq.~(\ref{cutoff}) gives a similar bound, $m_{\rm gelaton}^4 \le Tc_s^4$. These two are not identical since they correspond to different calculations; nevertheless, we see that consistency of the single field effective theory inevitably places an upper bound on the gelaton mass.

{\bf Conclusions:}
We have demonstrated that coupled multi-field inflationary models can have sound speeds less than unity
without the explicit introduction of higher powers of the kinetic term in the action. This occurs
when one of the fields is a gelaton, a heavy field that is tightly coupled to the inflaton, dressing its perturbations and reducing its sound speed. The resulting theory will exhibit equilateral type non-Gaussianity. This provides a possible mechanism
for UV-completion of $k$-inflation in the context of low energy supergravity models. Before horizon crossing, there will be a regime in which physical momenta $k/a$ are larger than $m_{\rm gelaton}$, so it is necessary to use the two field effective theory to describe the initial state of the system. This model provides an example of how non-standard initial conditions for the single field effective theory can arise from high energy physics. In particular the transition from the regime in which the two field theory is valid, to the gelaton regime where the single field theory is valid, could be sufficiently non-adiabatic to excited the state away from vacuum~\cite{prep}.

{\bf Acknowledgements:}
The authors thank C.~de Rham, N.~Barnaby, L.~Leblond, and S.~Shandera for discussions. Research at the Perimeter Institute is supported by the Government of Canada through Industry Canada and by the Province of Ontario through the Ministry of Research \& Innovation.

\end{document}